\documentclass[aps,prl,superscriptaddress,amsmath,nofootinbib,twocolumn,10pt]{revtex4-1}

\usepackage{color,hangcaption,hhline,psfrag,rotating,amssymb}
\usepackage[hang,nooneline]{subfigure}
\usepackage{dcolumn}
\usepackage{verbatim}
\usepackage{bm}

\begin{document}
\title{The shape of the $D \rightarrow K$ semileptonic 
form factor from full lattice QCD and $V_{cs}$}

\author{J.~Koponen}
\email[]{jonna.koponen@glasgow.ac.uk}
\author{C.~T.~H.~Davies}
\email[]{christine.davies@glasgow.ac.uk}
\author{G.~C.~Donald}
\affiliation{SUPA, School of Physics and Astronomy, University of Glasgow, Glasgow, G12 8QQ, UK}
\author{E. Follana}
\affiliation{Departamento de F\'{\i}sica Te\'{o}rica, Universidad de Zaragoza, E-50009 Zaragoza, Spain}
\author{G. P. Lepage}
\affiliation{Laboratory of Elementary-Particle Physics, Cornell University, Ithaca, New York 14853, USA}
\author{H. Na}
\affiliation{ALCF, Argonne National Laboratory, Argonne, IL 60439, USA}
\author{J.~Shigemitsu}
\affiliation{Physics Department, The Ohio State University, Columbus, Ohio 43210, USA}

\collaboration{HPQCD collaboration}
\homepage{http://www.physics.gla.ac.uk/HPQCD}
\noaffiliation

\date{\today}

\begin{abstract}
We present a new study of the form factors 
for $D \rightarrow K$ semileptonic decay from lattice QCD that 
allows us to compare the shape of the vector form factor 
to experiment and, for the first time, to extract $V_{cs}$ using 
results from all experimental $q^2$ bins. 
The valence quarks are implemented with the Highly Improved 
Staggered Quark action on MILC configurations that include 
$u$, $d$ and $s$ sea quarks. The scalar and vector currents 
are nonperturbatively normalised and, using phased boundary 
conditions, we are able to cover the full $q^2$ range 
accessible to experiment. 
Our result is $V_{cs}= 0.963(5)_{\mathrm{expt}}(14)_{\mathrm{lattice}}$. 
We also demonstrate that the form factors are insensitive 
to whether the spectator quark is $u/d$ or $s$, 
which has implications for other decay channels.  
\end{abstract}


\maketitle

{\it Introduction.} 
The analysis of weak semileptonic decays in 
which one meson changes into another 
and emits a $W$ boson allows direct determination of 
elements of the critical Cabibbo-Kobayashi-Maskawa (CKM) 
matrix of the Standard Model. At the same time we can test details 
of meson internal structure that are 
complementary to the results obtained 
by comparing QCD predictions of the meson mass or leptonic
decay constant to experiment. 
The information from 
QCD in semileptonic decay processes is parameterised by form 
factors that are functions of 
$q^2$, the square of the 4-momentum transfer from the initial 
to the final meson. Accurate calculations of the form factors in 
lattice QCD allow the $q^2$-dependence of the rate for such 
exclusive decays to be compared to experiment. The range of $q^2$ is
from $q^2_{max}$ 
where the daughter meson is at rest in the parent rest frame 
to $q^2=0$ where the daughter has maximum possible 
momentum in the opposite direction to the leptons from the virtual $W$. 
The appropriate CKM element is an overall factor 
in the comparison between lattice 
QCD and experiment and so can be determined. 
The accuracy achieved depends on 
the errors of the lattice QCD calculation but also 
the amount of experimental information that can 
be utilised in the comparison. Here, for the first time, we determine 
$V_{cs}$ from $D \to K \ell \nu$ decays using 
all the experimental $q^2$ bins, 
rather than 
just the $q^2 \rightarrow 0$ limit or total 
rate~\cite{fnaldk, Na:2010uf, etmdk}. 

$V_{cs}$ is the central element of the CKM matrix and 
a key one in tests of second row and second column 
unitarity. Unlike $V_{cd}$, which can be determined 
from neutrino and antineutrino interactions on 
the valence $d$ quarks in nuclei, the only 
direct determination methods 
capable of percent level accuracy for $V_{cs}$ 
are leptonic and semileptonic decays of charmed 
mesons. 
Here we obtain 
1.5\% accuracy for $V_{cs}$ using semileptonic decays, 
the best result to date, 
combining experimental information  
from BaBar~\cite{BaBar}, Belle~\cite{Belle}, BES~\cite{BESIII} and CLEO~\cite{CLEO}. 

This study also reveals the insensitivity of the charm
semileptonic form factors to the
mass of the spectator quark as it is varied between that of light
and strange. 

{\it Lattice Calculation.} 
For pseudoscalar 
to pseudoscalar meson decay only the vector piece of 
the weak current contributes. The
vector form factor, $f_+$, and the 
scalar form factor, $f_0$, appear in the matrix 
element as:
\begin{equation}\begin{split}
\langle K|V^\mu|D\rangle =
&f^{D\to K}_{+}(q^2)\bigg[p^\mu_D+p^\mu_K-\frac{M^2_D-M^2_K}{q^2}q^\mu\bigg]\\
+&f^{D\to K}_{0}(q^2)\frac{M^2_D-M^2_K}{q^2}q^\mu .
\label{eq:Jvector}
\end{split}\end{equation}
with the kinematic relation $f_+(0)=f_0(0)$. It is $f_+(q^2)$ 
that determines the experimental rate (since the contribution of 
$f_0$ is suppressed by the lepton mass) and is the form factor 
we concentrate on here. We have:
\begin{equation}
\frac{d\Gamma(D\rightarrow K\ell\nu)}{dq^2}=\frac{G^2_F|V_{cs}|^2}{24\pi^3}p^3|f_{+}(q^2)|^2 ,
\label{eq:rate}
\end{equation}
where $p$ is the 3-momentum of the $K$ in the $D$ rest-frame. 
We can separate $f_0$ and $f_+$ in Eq.~\ref{eq:Jvector} with a 
parallel calculation of $f_0$ from  
a scalar current matrix element. Using the partially conserved 
vector current (PCVC) relation ($\partial_{\mu}V_{\mu}=(m_c-m_s)S$) we have~\cite{Na:2010uf}:
\begin{equation}
\langle K|S|D\rangle = f^{D\to K}_{0}(q^2)\frac{M^2_D-M^2_K}{m_{c}-m_{s}},
\label{eq:Jscalar}
\end{equation}
where $m_c$ and $m_s$ are the quark masses in lattice QCD. 

\begin{figure}
\centering
\includegraphics[width=0.215\textwidth]{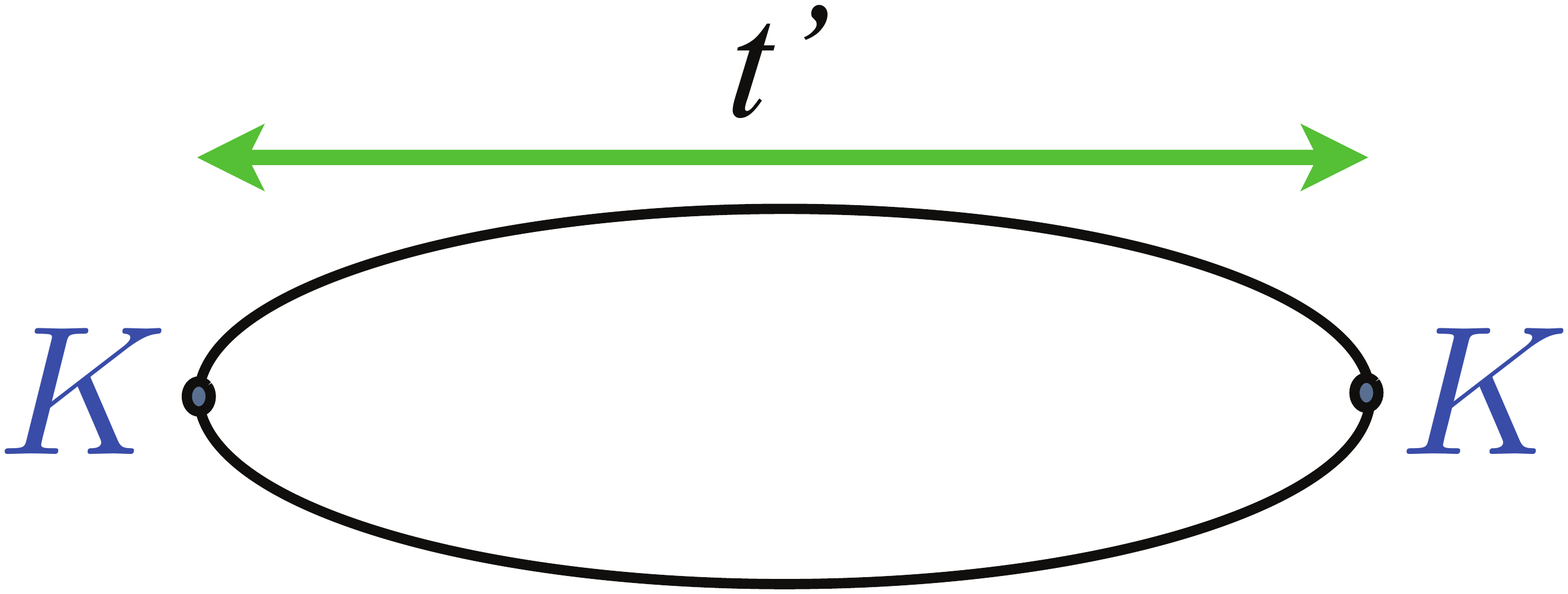}
\includegraphics[width=0.21\textwidth]{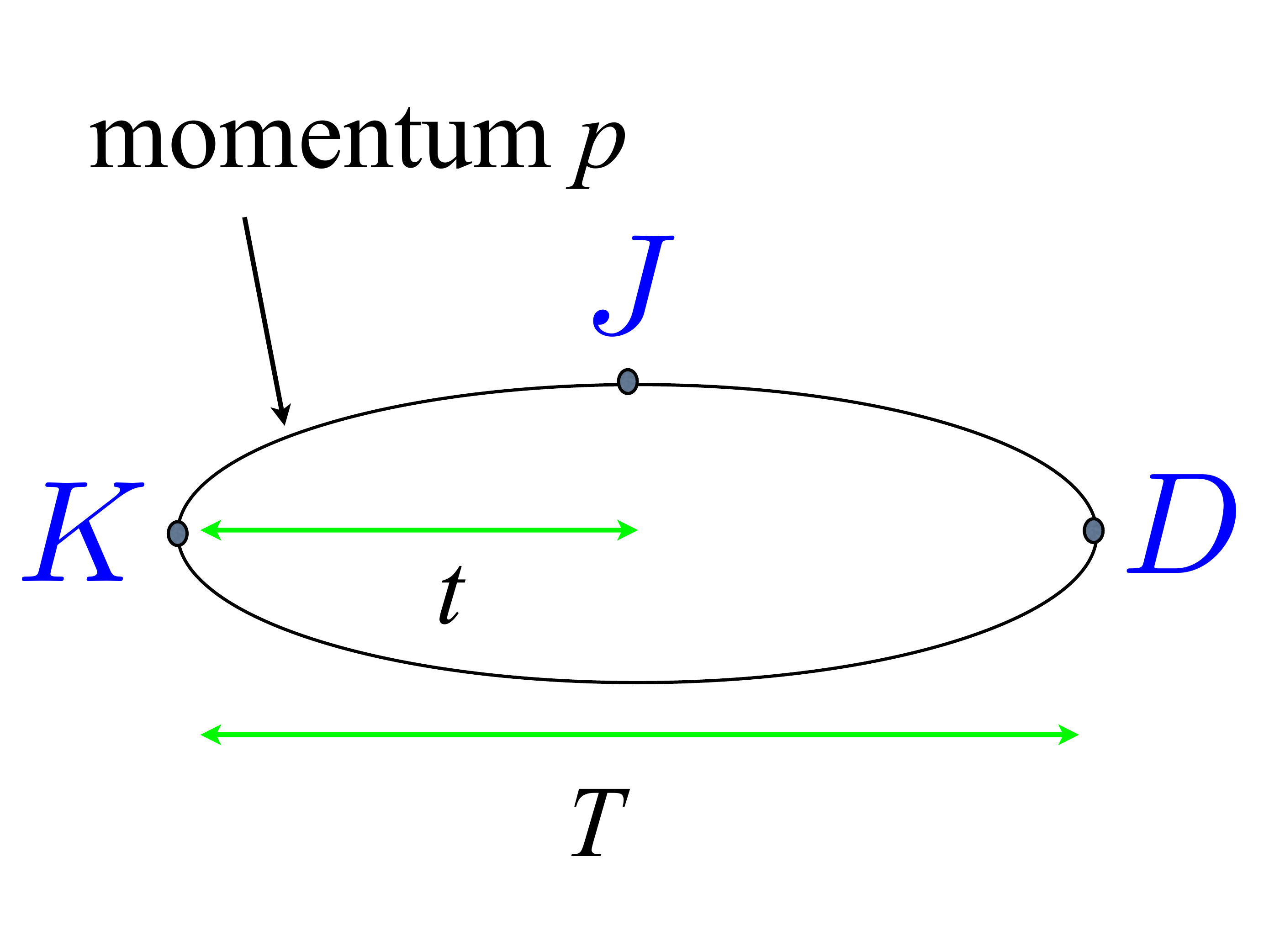}
\caption{2-point (left) and 3-point (right) correlators.}
\label{fig:2pt3ptdiagram}
\end{figure}

We use the Highly Improved Staggered 
Quark (HISQ) action~\cite{HISQ_PRD} for all the valence quarks. 
This action has very small discretisation errors, 
making it an excellent action for 
charm~\cite{HISQ_PRL, Dsdecayconst, jpsi} as well 
as lighter quarks. We calculate HISQ quark propagators from a local 
random wall source on 
gluon field configurations generated by the MILC 
collaboration that include the effect of 
$u$, $d$ and $s$ sea quarks using the asqtad 
formalism~\cite{MILCconfigs}. Table~\ref{tab:params} gives the 
parameters of the ensembles we use. By using multiple time sources 
on each configuration we generate very high statistics. 
The propagators are combined into meson correlation functions 
(2-point correlators) and correlation functions that allow 
for a $D$ to $K$ transition (3-point correlators). These are 
illustrated in Fig.~\ref{fig:2pt3ptdiagram}. We also use multiple 
values for $T$, the time separation between the sources of the 
two mesons, so that our fits can map out fully the $t$ and $T$ 
dependence of the 3-pt correlators for improved accuracy. 
In our $D \rightarrow K$ 3-pt correlators we keep the $D$ meson 
at rest but give the $K$ meson a non-zero momentum by using 
a `twisted' boundary condition~\cite{twist} on the $s$ quark propagator 
(see Fig.~\ref{fig:2pt3ptdiagram}). This enables us to map out 
the range of $q^2$ values. 

\begin{table}
\begin{tabular}{llllll}
\hline
\hline
Set &  $r_1/a$ & $au_0m_{l}^{asq}$ & $au_0m_{s}^{asq}$ & $m_l/m_{s,phys}$ & $L_s/a \times L_t/a$ \\
\hline
1 &  2.647(3) & 0.005 & 0.05 & 0.14 & 24 $\times$ 64 \\
2 &  2.618(3) & 0.01 & 0.05 & 0.29 & 20 $\times$ 64 \\
\hline
3 & 3.699(3) & 0.0062 & 0.031 & 0.24 & 28 $\times$ 96 \\
\hline 
\hline
Set & $am_{l}^{hisq}$  & $am_{s}^{hisq}$ & $am_{c}^{hisq}$ & $n_{cfg}\times n_t$ & $T$ \\
\hline
1 &  0.007 & 0.0489 & 0.622 & 2099 $\times$ 8(4) & 12, 15, 18  \\
2 &  0.0142 & 0.0496 & 0.63 & 2259 $\times$ 8 & 12, 15, 18  \\
\hline
3 & 0.008 & 0.0337 & 0.413 & 1911 $\times$ 4 & 16, 19, 20, 23 \\
\hline 
\hline
\end{tabular}
\caption{Upper table: Ensembles (sets) of MILC configurations used here. 
Sea 
(asqtad) quark masses $m_{l}^{asq}$ ($l = u/d$) and $m_s^{asq}$ 
use the MILC convention where $u_0$ is the plaquette 
tadpole parameter. 
The lattice spacing is given in units of $r_1$ after `smoothing'
~\cite{MILCconfigs}. We use $r_1=0.3133(23)$ fm~\cite{Davies:2009tsa}. 
Sets 1 and 2 are `coarse' ($a \approx 0.12$ fm) and set 3, 
`fine' ($a \approx 0.09$ fm).  The lattice size 
is given by $L_s^3 \times L_t$. Column 5 gives the 
sea light quark mass in units of the physical strange quark 
mass~\cite{Dsdecayconst} - the physical value for this 
ratio is 0.036~\cite{MILCconfigs}. 
Lower table: Valence $l$, $s$ and $c$ HISQ quark masses. 
The $l$ quark mass is tuned so that the $\pi$ mass is 
the same as that using the sea asqtad $l$ quarks~\cite{milc04}. 
The $s$ and $c$ quark masses are tuned to the 
physical values~\cite{Dsdecayconst}. 
We use $n_t$ time sources on each of the $n_{cfg}$ configurations. 
The final column lists the $T$ values used in the 3-pt 
correlators (see Fig.~\ref{fig:2pt3ptdiagram}). 
}
\label{tab:params}
\end{table}

The 2-point and 3-point correlators are fit using Bayesian methods~\cite{gplbayes} 
that allow us to include the effect of excited states, both 
`radial' excitations ($n$) and, because we are using staggered quarks, 
opposite parity mesons that give oscillating terms ($o$). 
We fit all the 2-point and 3-point correlators on a given ensemble 
at multiple momenta 
simultaneously to take account of correlations. The fit forms are: 
\begin{eqnarray}
C^{(P)}_{2pt} &=& \sum_{i_n,i_o} \{d^{(P)}_{i_n}\}^2 \mathrm{fn}(E^{(P)}_{i_n},t^{\prime}) - \{{\tilde{d}}^{(P)}_{i_o}\}^2 \mathrm{fo}(\tilde{E}^{(P)}_{i_o},t^{\prime}) \nonumber \\
C^{P\rightarrow Q}_{3pt}&=& \sum_{i_n,j_n} d^{(P)}_{i_n} \mathrm{fn}(E^{(P)}_{i_n},t) J^{nn}_{i_n,j_n} d^{(Q)}_{j_n} \mathrm{fn}(E^{(Q)}_{j_n},T-t) \nonumber \\
&-&\sum_{i_n,j_o} d^{(P)}_{i_n} \mathrm{fn}(E^{(P)}_{i_n},t) J^{no}_{i_n,j_o} \tilde{d}^{(Q)}_{j_o} \mathrm{fo}({\tilde{E}}^{(Q)}_{j_o},T-t) \nonumber \\
&+& ( n \leftrightarrow o)  
\label{eq:3ptfit}
\end{eqnarray}
with: 
\begin{eqnarray}
\mathrm{fn}(E,t) &=& e^{-Et} + e^{-E(L_t-t)} \nonumber \\
\mathrm{fo}(E,t) &=& (-1)^{t/a} \mathrm{fn}(E,t) 
\label{eq:fnfo}
\end{eqnarray}
Prior values and widths are taken as: ground-state, 2\% width; splitting 
between ground-state and excited energies, 600 MeV with 50\% width; splitting 
between ground-state and lowest oscillating state, 400 MeV for $D$ and 
350 MeV for $K$ with 50\% width; amplitudes, 0.01(1.0) for normal states 
and 0.01(0.5) for oscillating states; matrix elements, 0.01(2.0) for scalar 
currents and 0.01(1.0) for vector currents. We take the result from a 
5 exponential fit; $\chi^2/\mathrm{dof} <1$ and results and errors are stable there.  

\begin{figure}
\centering
\includegraphics[width=0.4\textwidth]{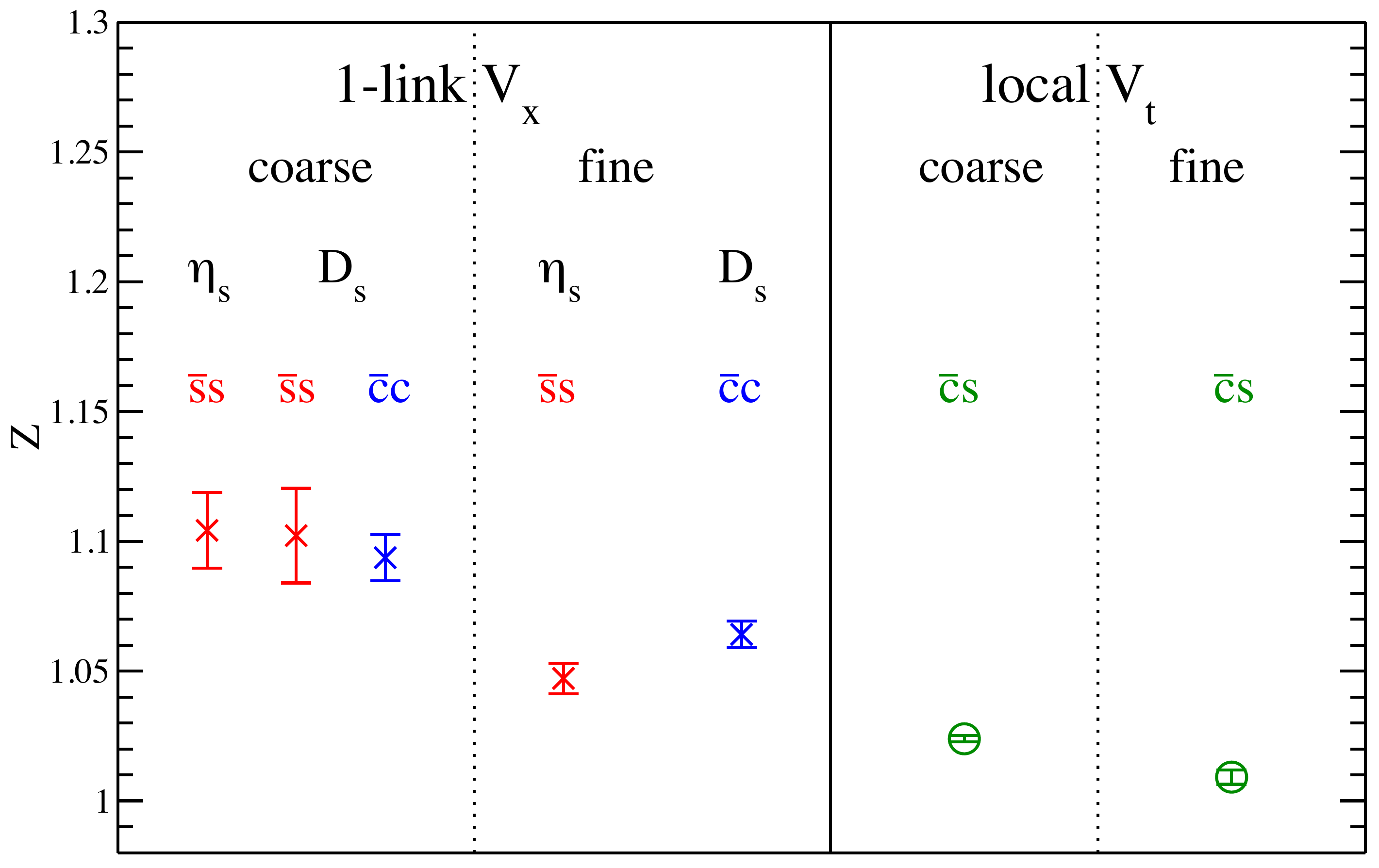}
\caption{Renormalization factors $Z$ for 1-link (left) and local 
vector currents determined on coarse set 2 and fine set 3.}
\label{fig:Z}
\end{figure}

$d_{i_n}$ in Eq.~\ref{eq:3ptfit} give the 
amplitudes for creation/annihilation of 
the $D/K$ mesons and $J_{i_n,j_n}$, the matrix elements of 
the vector/scalar current between $D$ and $K$. By matching to a continuum 
correlator with a relativistic normalisation of states and allowing 
for a renormalisation of the lattice current we see that the 
matrix elements between the ground state mesons that we want to determine
are given by: 
\begin{equation}
\langle D |J| K \rangle = Z\sqrt{4E^{(D)}_{0}E^{(K)}_{0}}J^{nn}_{0,0}.
\label{eq:me}
\end{equation}
The local scalar current that we use, when multiplied 
by the lattice quark mass, is absolutely normalised~\cite{Na:2010uf}, 
i.e. $f_0(q^2)$ can be extracted from eqs.~\ref{eq:Jscalar} and~\ref{eq:me}
setting $Z=1$.   
For the vector case we use 
two different currents: a point-split spatial vector current
(no gluon fields are included because we work in Coulomb gauge)
and a local temporal vector current. The spatial vector current is 
readily normalised for the $\overline{c}\gamma_ic$ 
and $\overline{s}\gamma_is$ cases 
by requiring that $Zf_+(0) = 1$. This is done in a calculation of the matrix 
element between two identical pseudoscalar mesons with the 
same non-zero momentum, 
achieved by giving a `twist' to the spectator quark~\cite{twist}.  
Fig.~\ref{fig:Z} shows the results of doing this on coarse 
set 2 and fine set 3. We see that the $Z$ factor is the 
same, to within few \% errors,  
for the $s$ and $c$ cases and is independent of the meson used
at source and sink of the 3-point correlator. 
We have also checked that results are independent of the momentum 
of the spectator quark and the sea quark masses (comparing sets 1 and 2). 
We therefore take the $Z$ factor for the 1-link spatial 
$\overline{c}\gamma_i s$ 
current to be that for the $\overline{c}\gamma_i c$ case.  
The local temporal vector $\overline{c}\gamma_0 s$ current is 
normalised by matching to the result for 
$f_0(q^2_{max})$ that 
we obtain from the absolutely normalised scalar current. 
This is done for $D_s$ decay to the $s\overline{s}$ pseudoscalar 
denoted $\eta_s$ (an unphysical state because it is not allowed to 
decay in lattice QCD).  
These $Z$ factors are also shown in Fig.~\ref{fig:Z}. 

Both the local scalar and the 1-link vector are `tasteless' currents 
in staggered quark parlance and so the 3-point correlator can be 
calculated between pseudoscalar mesons created using the 
local $\gamma_5$ (Goldstone) operator. 
The local temporal vector current has spin-taste $\gamma_0 \otimes \gamma_0$ 
and so, since tastes must cancel out in a 3-point correlator, 
it is used in a 3-point function between a charmed meson 
created with the local $\gamma_0\gamma_5$ operator 
and a Goldstone light meson. 
Using a different operator for the $D_{(s)}$ produces 
negligible effect here because the mass difference induced 
by taste-changing effects is very small 
(less than 4 MeV on coarse lattices and 1 MeV on fine)~\footnote{Taste-changing 
effects appear as 
an $\mathcal{O}(a^2)$ effect in the square of the mass for 
pseudoscalars. Differences in the mass itself are then suppressed by 
the mass for charmed mesons~\cite{HISQ_PRD}.}. 

\begin{table*}
\caption{
Results for form factors for $D \rightarrow K$ decay 
at 3 or 4 $q^2$ values per set corresponding 
to different $K$ momenta. 
}
\begin{ruledtabular}
\begin{tabular}{l|lll|lll|lll|ll}
Set & $q^2a^2$ & $f_+(q^2)$ & $f_0(q^2)$ & $q^2a^2$ & $f_+(q^2)$ & $f_0(q^2)$ & $q^2a^2$ & $f_+(q^2)$ & $f_0(q^2)$ & $q^2a^2$ & $f_0(q^2)$ \\  
\hline
1 & 0.010 & 0.755(13) & 0.753(14) & 0.43 & 1.090(8) & 0.896(5)  &  &  &  & 0.69 & 1.027(2) \\
\hline
2 & 0.002 & 0.751(8)  & 0.751(9)  & 0.34 & 0.994(5) & 0.862(3)  & 0.53 & 1.218(14) & 0.932(3) & 0.68 & 1.0186(15) \\
\hline
3 & 0.001 & 0.747(9)  & 0.746(9)  & 0.16 & 0.974(5) & 0.847(5)  & 0.26 & 1.200(14) & 0.948(6) & 0.34 & 1.011(2) \\
\end{tabular}
\end{ruledtabular}
\label{tab:ff}
\end{table*}

\begin{figure}
\centering
\includegraphics[width=0.4\textwidth]{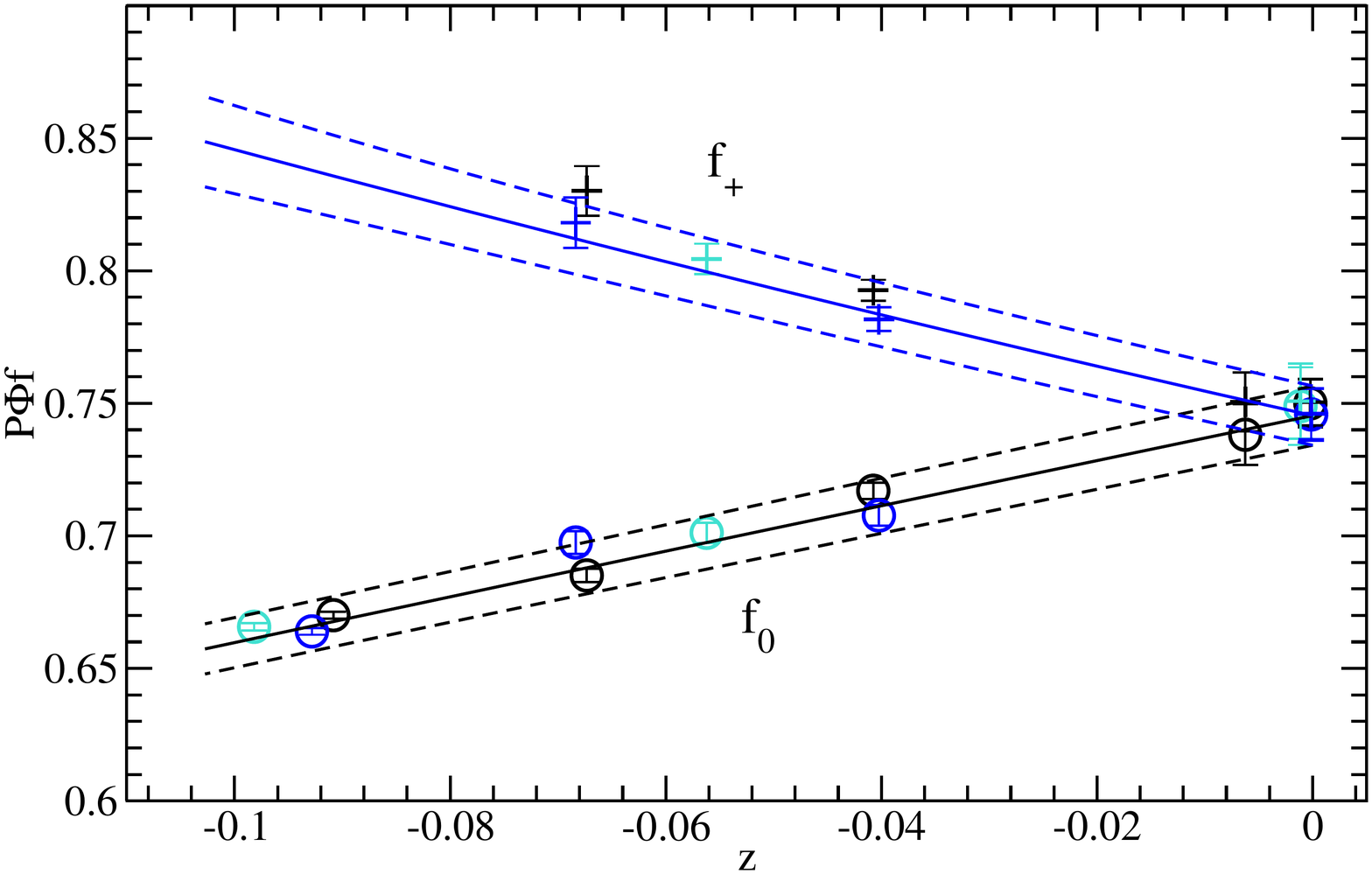}
\includegraphics[width=0.4\textwidth]{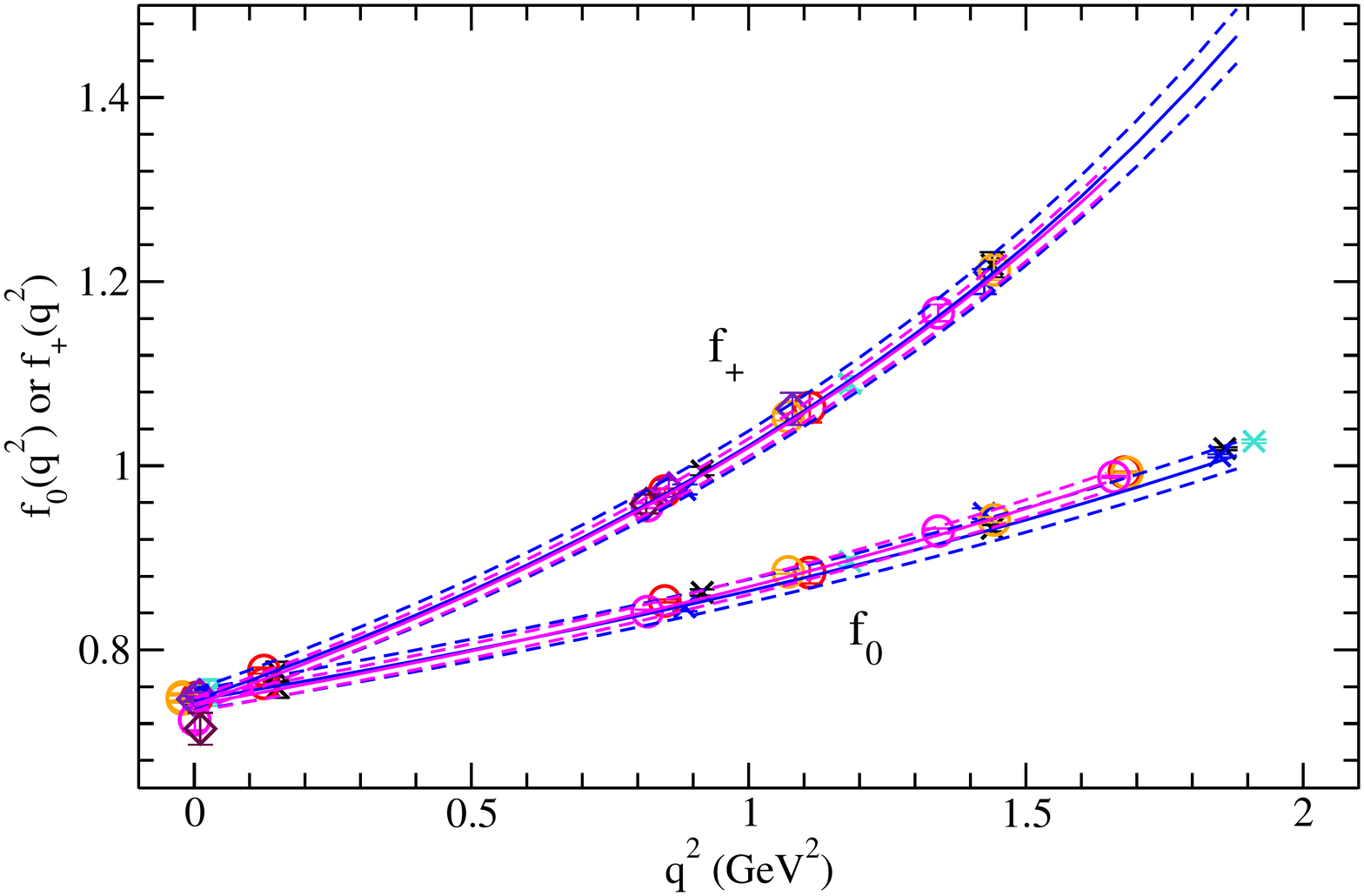}
\caption{Lattice results for $f_+$ and $f_0$ in (upper plot) $z$-space 
and (lower plot) $q^2$-space. Upper plot shows $D \rightarrow K$ 
$f_+$ (plus signs) and $f_0$ (circles); 
set 1 (light blue), set 2 (black) and set 3 (dark blue). 
Our fit (in the $a \rightarrow 0$ and $m_l \rightarrow m_{l,phys}$ limit) 
is shown with solid and dashed lines. 
The lower plot shows $f_+$ and $f_0$ for 
$D \rightarrow K$ (crosses) and $D_s \rightarrow \eta_s$ (circles 
for 1-link vector and diamonds for local temporal vector currents).  
The results from the z-space fits are plotted with lines - blue for 
$D \rightarrow K$ and pink for $D_s \rightarrow \eta_s$. }
\label{fig:DKzfit}
\end{figure}

{\it Results.} 
Table~\ref{tab:ff} gives our raw results for $f_+$ and 
$f_0$ for $D \rightarrow K$ from combining (spatial) vector and 
scalar matrix elements, after renormalising the vector. 
To determine the functional shape of the form factors we 
transform to $z$-space where: 
\begin{equation}
z=\frac{\sqrt{t_+-q^2}-\sqrt{t_+-t_0}}{\sqrt{t_+-q^2}+\sqrt{t_+-t_0}},\quad t_{\pm}=(m_D\pm m_K)^2 .
\label{eq:zspace}
\end{equation}
This maps the semi-leptonic region, $0<q^2<t_{\_}$ to the interior 
of the unit circle, allowing for polynomial fits in $z$.  
We then fit the form factors to 
\begin{equation}
f(q^2)=\frac{1}{P(q^2)\Phi(q^2)}\sum_{n=0}^N b_n z^n.
\label{eq:ffit}\end{equation}
To combine fits for $f_+$ and $f_0$ it is convenient for us 
to take $t_0=0$ (so that $q^2=0$ maps to $z=0$) and 
to take the simplest form~\cite{Bourrely:2008za} 
for the product $P(q^2)\Phi(q^2)$, 
which is
$(1-q^2/M_X^2)$ where $M_X$ is the appropriate pole mass, 
$M_{D_s^*}$ for $f_+$ and $M_{D_{s0}^*}$ for $f_0$. 

Fig.~\ref{fig:DKzfit} shows our results for $P\Phi\times f$ in
$z$-space, where it is clear they have a very simple form. 
To obtain results in the continuum and physical light quark 
mass limits, we allow for dependence of the coefficients $b_n$ in 
Eq.~\ref{eq:ffit} on $a$ and valence and sea $m_{l}$ 
(using chiral parameter $\delta_l = 0.25m_l/m_{s,phys}$ from Table~\ref{tab:params}) as: 
\begin{eqnarray}
b_n(a, m_l) =A_n\{1&+&B_n a^2 + C_n a^4 + D_n\delta_l \nonumber \\ 
&+&E_n(\delta_{l}\ln [\delta_{l}] + F_na^2\delta_{l})\}
\label{eq:bnfit}
\end{eqnarray}
Priors are taken as: $A_0$: 0.750(75), $A_n$, $n>0$: 0.0(2.0),
$B_n$: 0.0(3), $C_n$: 0.0(1.0), $D_n$: 0.0(5), $E_n$, $F_n$: 0.0(1.0).
We include coefficients up to $n=4$, with a constraint on the 
$n=4$ value~\cite{Bourrely:2008za}. 
Coefficients are independent for $f_0$ and $f_+$ except for 
the kinematic constraint that $b_0$ should be the same for both. 
From the fits we extract $b_{n,phys}=b_n(a=0, m_l=m_{l,phys})$. 

\begin{figure}
\centering
\includegraphics[width=0.45\textwidth]{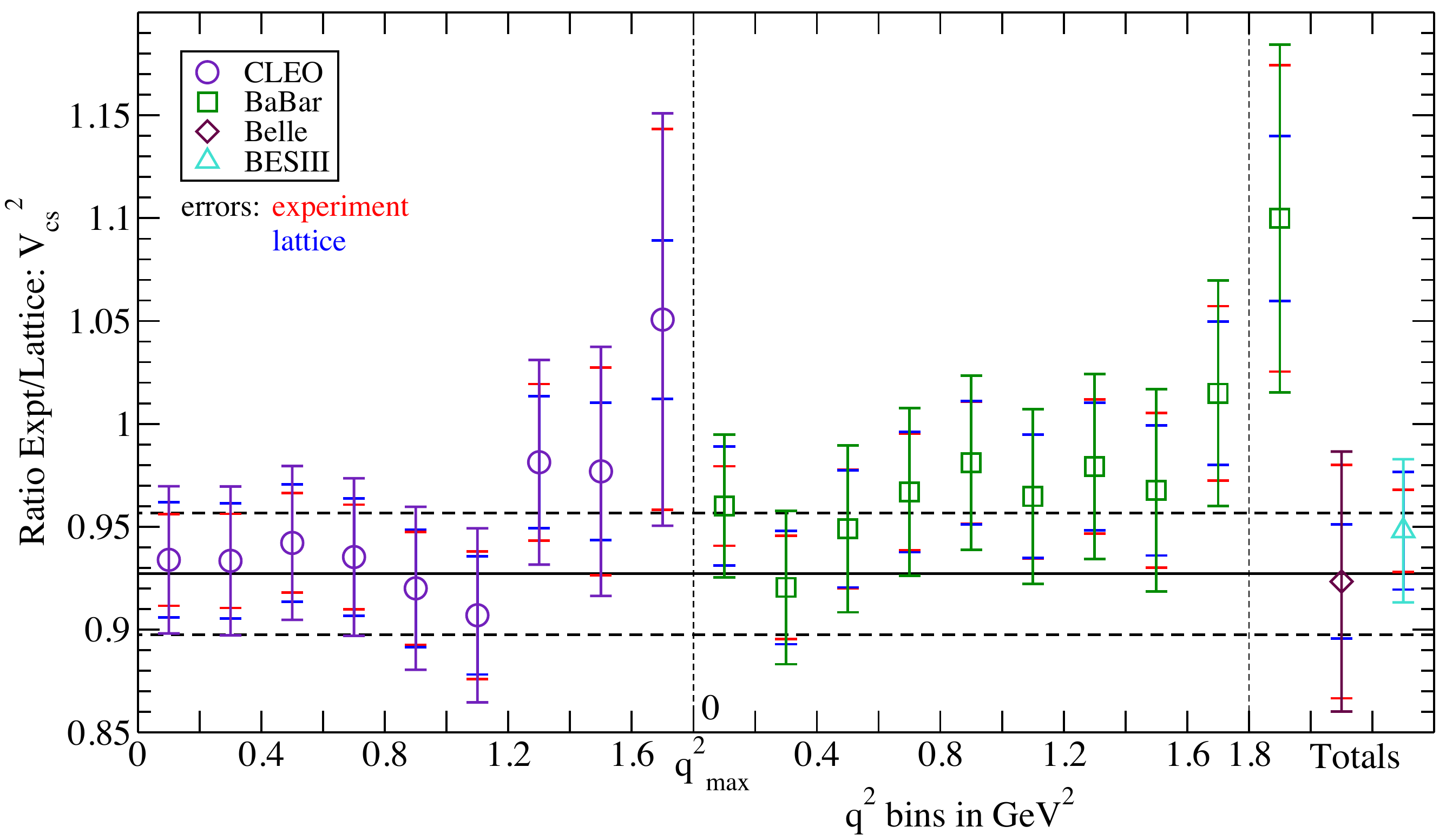}
\caption{Ratio of experimental to lattice results in each $q^2$
bin for $D^0\to K^-\ell^+\nu$, using CLEO~\cite{CLEO} and 
BaBar~\cite{BaBar} data. The last 3 bins are total rates for 
BaBar~\cite{BaBar}, Belle~\cite{Belle} and BESIII~\cite{BESIII}. 
Error bars from experiment and from lattice QCD are marked 
separately on each point. The horizontal lines give our fitted 
result for $V_{cs}^2$ with its error.  
}
\label{fig:compexp}
\end{figure}

Our physical curve in $z$-space is converted back to 
$q^2$ space giving the lower plot of Fig.~\ref{fig:DKzfit}. 
We integrate
the factor $p^3|f_+(q^2)|^2$ from Eq.~\ref{eq:rate} over 
the experimental bins 
in $q^2$ (the same for CLEO and BaBar) and can then make a bin-by-bin 
comparison, including the correlations between bins for lattice 
QCD and experiment. This comparison is shown in Fig.~\ref{fig:compexp} 
in which we plot the ratio of experiment to lattice QCD for 
each bin, which is a value for $|V_{cs}|^2$ from that bin. We also show 
the result of fitting a weighted average over the bins to 
obtain a final value for $|V_{cs}|$. We use  
CLEO~\cite{CLEO} and BaBar~\cite{BaBar} binned data and
 BaBar, Belle~\cite{Belle} and BESIII~\cite{BESIII} 
total rates for $D^0 \rightarrow K^-$ to obtain 
$|V_{cs}|=0.963(5)_{\mathrm{expt}}(14)_{\mathrm{lattice}}$. 
Different subsets of experimental 
results give consistent values; the error is smallest
using all of them. For the binned data the experimental 
results are most accurate at low $q^2$, the lattice QCD results, 
at high $q^2$. The optimal bins for the combination are 
1 to 6 ($q^2 =$ 0-1.2 $\mathrm{GeV}^2$), see Fig.~\ref{fig:compexp}. 

\begin{figure}
\centering
\includegraphics[width=0.45\textwidth]{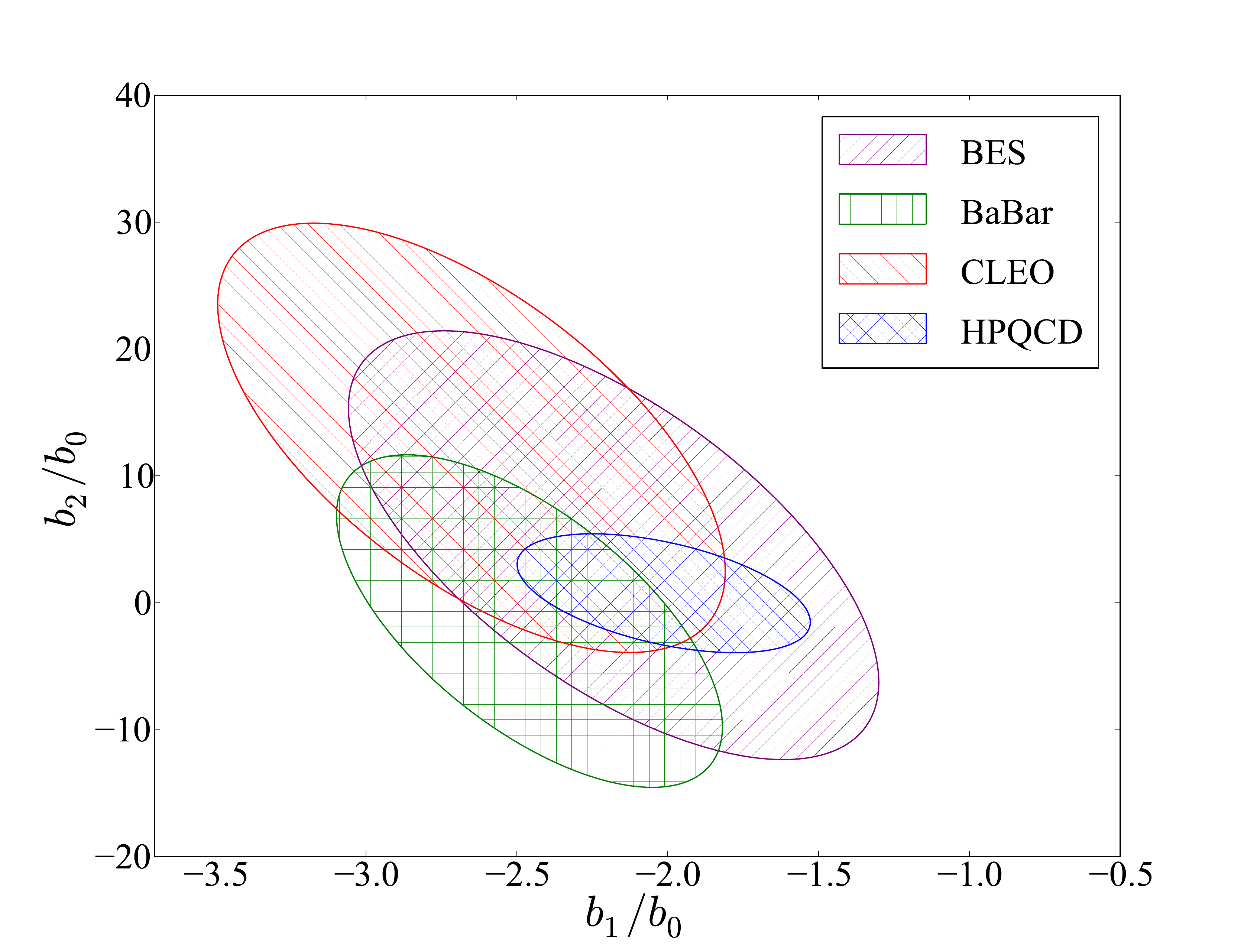}
\caption{68\% confidence limits on the shape parameter ratios 
$b_1/b_0$ and $b_2/b_0$ from a 3-parameter $z$-space fit 
to $f_+$ (Eq.~\ref{eq:ffit} using $P\Phi$ and 
$t_0$ from~\cite{CLEO}), for lattice QCD 
and experiment~\cite{CLEO, BaBar, BESIII, private}. 
BaBar parameters shown are from 
our fit to the binned correlated data. 
Our results are: $b_1/b_0=-2.01(23)$, $b_2/b_0=0.75(2.5)$ and correlation, 
$\rho=-0.56$. 
}
\label{fig:ellipse}
\end{figure}

We can also compare the shape more accurately to experiment using 
a common $z$-space expansion. We take $t_0 =t_+(1-(1-t_-/t_+0)^{1/2})$ 
in Eq.~\ref{eq:ffit} and a specific 
form for $P(q^2)\Phi(q^2)$ given in~\cite{CLEO, Hill:2006ub}. 
Fig.~\ref{fig:ellipse} compares our results at the physical 
point for $b_1/b_0$ 
and $b_2/b_0$ to experiment for this case.  
The agreement is excellent. 

Finally, we note that Fig.~\ref{fig:DKzfit} shows both the 
$D \rightarrow K$ and $D_s \rightarrow \eta_s$ form factors 
as a function of $q^2$. 
The two processes differ in their 
spectator quark - $D \rightarrow K$ has a $u/d$ spectator 
and $D_s \rightarrow \eta_s$ an $s$ - but their form factors agree to 2\%.   
This was also found for 
$B_{(s)}\rightarrow D_{(s)}$ decays in~\cite{Bailey:2012rr} and is likely to be a generic feature of 
heavy quark decays. Model calculations give 
varying results~\cite{richman, isgw2} with $\mathcal{O}$(10\%) 
effects possible. 

Fig.~\ref{fig:DKzfit} also demonstrates how small discretisation 
errors are with results from coarse and fine lattices lying on 
top of each other. A further check of this is a comparison 
of the $D_s \rightarrow \eta_s$ 
form factors from 1-link spatial and local temporal vector currents 
which also show no difference. 

{\it Conclusions.} 
We have calculated the form factors for $D \to K$ 
semileptonic decay from full lattice QCD, and compared the shape of the 
vector form factor $f_+(q^2)$ to experiment across the full $q^2$ range. 
We extract $V_{cs}$ for the 
first time using all $q^2$ bins. Our result is 
$V_{cs}=0.963(5)_{\text{exp}}(14)_{\text{lattice}}$, which improves the 
accuracy of our previous world's best determination~\cite{Na:2010uf} 
of $V_{cs}$ by over 50\%. At $q^2=0$ we obtain $f_+(0)=0.745(11)$.   

Our result for $V_{cs}$ agrees with that from CKM matrix 
unitarity (0.97344(16)~\cite{pdg}) and gives 
separate tests of the second row and column that agree 
with unitarity to 3\%. 
Combining 
the $D_s$ leptonic decay rate with lattice QCD results for the 
$D_s$ decay constant~\cite{Dsdecayconst} yields a higher but 
consistent 
$V_{cs}$, for example $1.001(10)_{\mathrm{latt}}(26)_{\mathrm{expt}}$ 
using recent Belle 
results~\cite{bellefds}. 

We see no difference between form factors for a $s$ or $u/d$ 
spectator quark to the $c \rightarrow s$ decay. This is also 
true for $c \rightarrow d$ decays comparing $D \rightarrow \pi \ell \nu$ and 
$D_s \rightarrow K \ell \nu$. These results will be discussed elsewhere. 

{\it Acknowledgements.} We are grateful to MILC for the use of their 
gauge configurations and to Lawrence Gibbons 
for useful discussions. 
We used the Darwin Supercomputer 
as part of the DiRAC facility jointly
funded by STFC, BIS 
and the Universities of Cambridge and Glasgow. 
This work was funded by STFC, the Royal Society and the 
Wolfson Foundation, MICINN, EU ITN StrongNet, NSF and DOE.

\bibliography{dk}

\end{document}